\documentclass[a4paper,conference]{IEEEtran}

\usepackage[utf8]{inputenc}

\usepackage{color}

\usepackage{amsmath,amssymb,bm}
\usepackage{cite}
\usepackage[caption=false,font=footnotesize]{subfig}

\usepackage[acronyms]{glossaries}

\usepackage[export]{adjustbox}

\usepackage[hidelinks]{hyperref}

\newacronym{uwb}{UWB}{ultra-wideband}
\newacronym{cir}{CIR}{channel impulse response}
\newacronym{cnn}{CNN}{convolutional neural network}
\newacronym{fft}{FFT}{fast Fourier transform}
\newacronym{fmcw}{FMCW}{frequency-modulated continuous-wave}
\newacronym{snr}{SNR}{signal-to-noise ratio}
\newacronym{auc}{AUC}{area under the receiver operating curve}
\newacronym{vmp}{VMP}{variational message passing}
\newacronym{flop}{FLOP}{floating point operation}
\newcommand{\datasetname}{UWBCarGraz}

\newcommand{\nn}{\nonumber}
\newcommand{\ist}{\hspace*{.3mm}}

\newcommand{\iist}{\hspace*{1mm}}

\newcommand{\T}{\text{T}}

\IEEEoverridecommandlockouts

\begin{document}
	\author{Jakob M\"oderl, Stefan Posch, Franz Pernkopf and Klaus Witrisal\\Graz University of Technology (jakob.moederl@tugraz.at)%
		\thanks{
			This research was partly funded by the Austrian Research PromotionAgency (FFG) within the project SEAMAL Front (project number: 880598). Furthermore, the financial support by the Christian Doppler Research Association, the Austrian Federal Ministry for Digital and Economic Affairs and the National Foundation for Research, Technology and Development is gratefully acknowledged.
			
			All authors are associated with Graz University of Technology.
			Stefan Posch and Klaus Witrisal are further associated with the Christian Doppler Laboratory for Location-aware Electronic Systems.}}
	\title{\Huge ``\datasetname'' Dataset for Car Occupancy Detection using Ultra-Wideband Radar}
	\maketitle
	
	\begin{abstract}
		We present a data-driven car occupancy detection algorithm using \acrlong{uwb} radar based on the ResNet architecture. The algorithm is trained on a dataset of \acrlongpl{cir} obtained from measurements at three different activity levels of the occupants (i.e. breathing, talking, moving).
		We compare the presented algorithm against a state-of-the-art car occupancy detection algorithm based on \gls{vmp}.
		Our presented ResNet architecture is able to outperform the \gls{vmp} algorithm in terms of the \gls{auc} at low \glspl{snr} for all three activity levels of the target. Specifically, for an \gls{snr} of $-20\,\text{dB}$ the \gls{vmp} detector achieves an \gls{auc} of $0.87$ while the ResNet architecture achieves an \gls{auc} of $0.91$ if the target is sitting still and breathing naturally. The difference in performance for the other activities is similar.
		To facilitate the implementation in the onboard computer of a car we perform an ablation study to optimize the tradeoff between performance and computational complexity for several ResNet architectures.
		The dataset used to train and evaluate the algorithm is openly accessible. This facilitates an easy comparison in future works.
	\end{abstract}
	\glsresetall
	
	\section{Introduction}
	Radar will become a pervasive technology through the advent of civilian applications e.g. in assisted living for vital-sign monitoring \cite{paternianiIEEEProc2023:vital-signs} and human activity recognition \cite{yangBHI2023:human-activity-radar-challenge}.
	Similarly, radar can be deployed in the passenger cabin of a car, e.g. to detect occupants. Future cars will be required to detect whether they are empty or occupied when they are locked to prevent the confinement (and resulting endangerment) of small children \cite{euroncapRoadmap2025,qinyi2020,aghaei2016}.
	New generation car models are equipped with \gls{uwb} nodes for keyless car access, that can be used for opportunistic sensing of occupants \cite{moederlEuRad2022,moederlICASSP2023,ma20}.
	The signal received by the radar contains the radar response from the target, but also includes additive noise and (strong) clutter \cite{moederlEuRad2022,moederlICASSP2023}.
	To distinguish the target signal from clutter, most works focus on detecting variations of the \gls{cir} over time, which correspond to movements of the target \cite{schleicher13,leibEuCAP2010:vital-sign-uwb-correlation,lazaroAccess2021,munteSensors2022}.
	This movement can range from a subtle chest movement due to the respiration of the target to rather larger movements, such as limb movements or a repositioning of the torso.
	
	Several model-driven \cite{baird17,lazaroAccess2021,munteSensors2022,songMDPIEntropy2021,diewald16} and data-driven \cite{alizadeh19:IEEEsensors,srirangaSSCI2022,ma20,numanSensor2023,kwonIRS202} approaches to detect car occupants can be found in the literature which utilize either \gls{uwb} or \gls{fmcw} radar at various carrier frequencies.
	However, none of these works have made the dataset used to train and/or evaluate their algorithms publicly available. This makes it difficult to compare and assess the results obtained by different methods in a fair and unbiased manner.
	This is exacerbated for data-driven methods, which can not be properly reproduced and validated without the data used to train the network.

	In previous works, we focused on modeling and detecting the case of small movement of the target due to respiration \cite{moederlEuRad2022,moederlICASSP2023}.
	For such small movements, the \glspl{cir} received from the target over time can be approximately factorized into a time-independent channel multiplied with the time-varying movement amplitude of the target.
	In \cite{moederlICASSP2023}, \gls{vmp} is used to estimate the factors of this product and thereby detect the target.
	The resulting detector was shown to outperform other model-based approaches such as the estimator-correlator \cite{moederlEuRad2022}, \gls{fft}-based detectors or windowed-energy detectors \cite{kilic14} on simulated and real data \cite{moederlICASSP2023}.
	However, this signal model is only valid for small body movements, but not for larger movements such as limb or torso movements.
	Furthermore, the prior chosen to model the respiratory motion is rather general and might not represent the true physiological motion accurately, e.g. when the target is speaking.
	In this paper, we address these issues using a data-driven approach, where the statistical relations of the detection problem are learned by a ResNet architecture \cite{heCVPR2016:ResNet}.
    Hence, the two main contributions of this paper are
    \begin{itemize}
    \item We design and train a computationally efficient ResNet architecture for car occupancy detection and show that it outperforms state-of-the art model-based methods, such as the \gls{vmp} detector \cite{moederlICASSP2023}, for car occupancy detection in terms of the \gls{auc} at low \glspl{snr}.
	\item We present the ``\datasetname'' dataset, an open dataset for car occupancy detection using \gls{uwb} radar, available at \cite{moederl2023:Dataset}.
    \end{itemize}

	\section{System Model}
	
	We consider a radar system which repeatedly transmits a pulse $s(\tau)$ with center frequency $f_{\text{c}}$ and bandwidth $B$ at times $t=mT_{\text{st}}$, $m\in\{0,1,\cdots,M-1\}$ with repetition interval $T_{\text{st}}$, where $\tau$ denotes ``fast time'' with a magnitude of nanoseconds and $t$ denotes ``slow time'' with a magnitude of seconds.
	The signal propagates through the environment where it is reflected by the target as well as all other objects in the environment.
	At each time instance $t=mT_{\text{st}}$, $N$ samples with sampling interval $T_{\text{ft}}$ are obtained.
	Assuming that $T_{\text{st}}\gg N T_{\text{ft}}$ is chosen large enough such that successive repetitions of the transmit signal do not interfere with each other, the received signal $r_m[n]$, $n\in\{0,\ist 1,\ist \cdots, \ist N-1\}$ at repetition $m$ is modeled as
	\begin{equation}\label{eq:received-signal-general}
		r_m[n] = \big(h(\cdot, mT_{\text{st}}) \ast s(\cdot)\big)(nT_{\text{ft}}) + v(nT_{\text{ft}},mT_{\text{st}})
	\end{equation}
	where $\big(h(\cdot)\ast s(\cdot)\big)(nT_{\text{ft}})$ denotes the (fast time) convolution of the functions $h$ and $s$ evaluated at $nT_{\text{ft}}$, $h(\tau, t)$ is the time-varying \gls{cir} and $v(\tau,t)$ is an additive noise process which is assumed to be white and independent across different repetitions $m$.
    We use a multipath propagation model \cite{michelusi12:part1,molischBookWirelessComm} to describe the \gls{cir} $h(\tau,t)$ as a sum of $L$ multipath components. Of these $L$ multipath components, $K$ components interact with the target and $L-K$ components interact only with the environment. We assume a static environment and express the \gls{cir} $h(\tau, t)$ as
	\begin{align}\label{eq:multipath-cir-model}
		h(\tau,t) = &\sum_{k=1}^{K} \alpha_k(t) \delta(\tau - \tau_k(t)) + \sum_{k=K+1}^{L} \alpha_k \delta(\tau - \tau_k)
	\end{align}
	where $\delta(\cdot)$ denotes the Dirac-delta function, $\alpha_k$ and $\tau_k$ denote the amplitude and delay of the $k$-th multipath component, which are time-varying for $k\leq K$, i.e. for all multipath components associated with the target but not for the multipath components associated with the static environment.
	Inserting \eqref{eq:multipath-cir-model} into \eqref{eq:received-signal-general} we get
	\begin{align}\label{eq:received-signal-multipath}
		r_m[n] =& \sum_{k=1}^{K} \alpha_k(mT_{\text{st}}) s(nT_{\text{ft}}-\tau_k(mT_{\text{st}})) \nn \\
		 &\quad + \sum_{k=K+1}^{L}\alpha_k s(nT_{\text{ft}}-\tau_k) + v(nT_{\text{ft}}, mT_{\text{st}})
		 .
	\end{align}
	
	Let $\bm{r}_m = \big[r_m[0]\ist\iist r_m[1]\ist \cdots \ist r_m[N-1]\big]^\T$ denote the vector of the received signal samples at time $m$.
	We stack the signals received at different times $m$ into a matrix $\bm{R}\in\mathbb{C}^{N\times M}$
	\begin{equation}
		\bm{R} = \big[\bm{r}_0\ist\iist \bm{r}_1\ist \cdots \ist \bm{r}_{M-1}\big]
		.
	\end{equation}
	From \eqref{eq:received-signal-multipath} it follows that the columns $\bm{r}_m$ of $\bm{R}$ can be decomposed into a constant part $\bar{\bm{r}} = \frac{1}{M}\sum_{m=1}^{M}\bm{r}_m$ and a time-varying part $\tilde{\bm{r}}_m = \bm{r}_m-\bar{\bm{r}}$ which includes the variations in the \gls{cir} over time due to the target plus noise.
	If no target is present, the columns $\tilde{\bm{r}}_m$ will contain only noise.
	Hence, we focus on detecting if the mean-removed received signal
	\begin{equation}
		\tilde{\bm{R}} = [\tilde{\bm{r}}_0\ist \iist \tilde{\bm{r}}_1 \ist \cdots \ist \tilde{\bm{r}}_{M-1}]
	\end{equation}
	contains only white noise or, if variations due to a target are present.\footnote{Note, that also the static part $\bar{\bm{r}}$ potentially depends on the target.
	However, to avoid false positives due to e.g. a suitcase put in the car we focus solely on the time-varying part $\tilde{\bm{R}}$ in the detection process.}

	\section{\datasetname\  Dataset Description}

\begin{figure}
		\centering
		\includegraphics{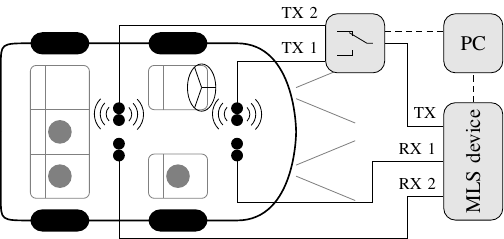}
		\caption{Illustration of the measurement setup. The participant positions are indicated with gray circles.}
		\label{fig:meas:setup}
	\end{figure}

	A measurement campaign was conducted at Graz University of Technology with 34 adult participants (6 female, 28 male).
	The participants are volunteers recruited from the staff and students of the university.
	To gather \glspl{cir}, a maximum length-sequence (MLS) correlative channel sounder \cite{sachsICUW2007:M-sequence} with a nominal frequency range of $3.8$--$10.2\,\text{GHz}$ was used to measure the \gls{cir} between one transmit and two receive antennas with a repetition interval of $T_{\text{st}}=0.1\,\text{s}$.
	Two pairs of antennas were placed under the ceiling of the passenger cabin near the A- and B-pillars.
	One antenna of each pair was receiving while the transmit signal was connected to an RF-switch such that either of the remaining two antennas could be used as transmit antenna as shown in Figure \ref{fig:meas:setup}.
	The equipment, including channel sounder, cables and connectors, was calibrated before the measurement except for the antennas.
	Dipole antennas with circular patches have been used \cite[Figure B.5b]{krall08:PhD}.
	The pairs of antennas were placed such that the zeros in the radiation patterns of the antennas pointed towards each other in order to minimize the mutual coupling between them.
	In the post processing, the data was multiplied in the frequency domain with a raised cosine pulse with center frequency $f_{\text{c}}=6.5\,\text{GHz}$, bandwidth $B=500\,\text{MHz}$ and roll-off factor of $\beta=0.5$, corresponding to the \gls{uwb} Channel 5 in \cite{IEEE:802.15.4}.
	A second variant of the dataset with a bandwidth of $5\,\text{GHz}$ centered around $f_{\text{c}}=7\,\text{GHz}$ is also available.
	Additionally, the expansion of the chest in each breathing cycle of the participants was recorded with an elastic belt \cite{respirationBelt}.
	Therefore, the dataset can be used for breathing motion estimation as well.
	For more details about the calibration and data pre-processing we refer the reader to the technical document accompanying the dataset \cite{moederl2023:Dataset}.
	
	The participants were seated in the front or rear seat on the passenger side or in the rear middle seat as shown in Figure \ref{fig:meas:setup}.
	The participants were not seated on the driver side for practical reasons.
	However, due to the symmetry of the setup we do not expect the \glspl{cir} recorded with people sitting in the driver side to differ substantially from the ones recorded with people on the passenger seats.
	In order to record different types of movements, the participants were asked to perform three different activities for approximately 2 minutes each.
	The data from each activity was then segmented into non-overlapping samples with $10\,\text{s}$ length.
	The activities are
	\begin{itemize}
		\item[(i)] breathing: sitting still without intentional movements while breathing naturally,
		\item[(ii)] talking: speaking out loud either a freely improvised monologue or by reading from a book, and
		\item[(iii)] moving: looking around and searching for small objects hidden inside the car.
	\end{itemize}
	Additionally, 5 minutes of data of the empty car was also recorded.
	Measurements have been performed with two different cars: a hatchback compact car (Car 1, Seat Leon) and a minivan (Car 2, Citro\"en C4 Picasso).
	Out of the 34 participants, 9 were recorded twice, once in each car, while the remaining 25 participants were recorded only in one of the cars for a total of 21 participants in the hatchback compact car and 22 participants in the minivan.
	Table \ref{tab:meas:bio-metrics} list the min, max and median of the participants height, weight and age. It is evident from Table \ref{tab:meas:bio-metrics}, that a diverse group participated in the measurement campaign.
	\begin{table}
		\centering
		\caption{Bio-metrics of the participants.}
		\label{tab:meas:bio-metrics}
		\begin{tabular}{r|cccc}
			\hline\hline
			& min & max & median & unit \\ \hline
			height & 158 & 190 & 178.5 & cm \\
			weight & 45 & 158 & 72 & kg \\
			age & 22 & 64 & 29	& years	\\
			\hline \hline	
		\end{tabular}
	\end{table}

	\begin{figure*}
		\centering
		\subfloat[]{\includegraphics{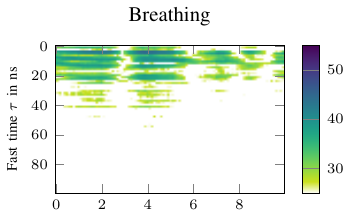}}~\subfloat[]{\includegraphics{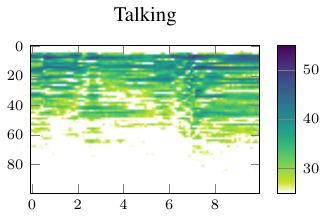}}~\subfloat[]{\includegraphics{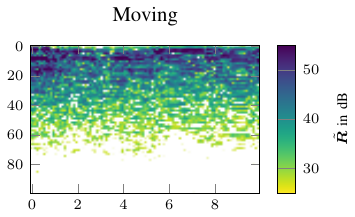}}\\
		\subfloat[]{\includegraphics{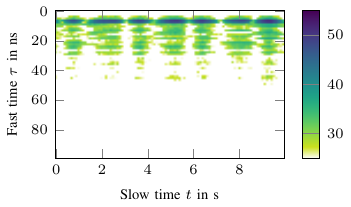}}~\subfloat[]{\includegraphics{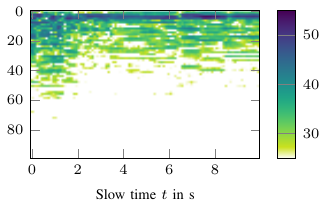}}~\subfloat[]{\includegraphics{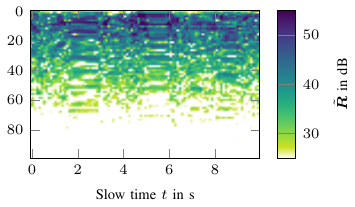}}\\
%		\subfloat[]{\input{./pgf/dataset-example-7.tex}}
		\caption{Example signals from the dataset. The plots show the magnitude of the received signal $\tilde{\bm{R}}$ in dB for the breathing activity (a, d), talking activity (b, e) and moving activity (c, f). The x-axis is the measurement time in seconds (``slow time'') whereas the y-axis is the propagation delay in nanoseconds (``fast time''). Note, that the color scale is clipped so that all values $\geq 55\,\text{dB}$ map to dark blue.}
		\label{fig:dataset:examples}
	\end{figure*}

	Examples of the time-varying part of the received signal $\tilde{\bm{R}}$ of the three different activity levels as well as for the empty car are shown in Figure \ref{fig:dataset:examples}.
	Each column of $\tilde{\bm{R}}$ corresponds to the time-varying part of the \gls{cir} at time $t=nT_{\text{st}}$ and each row corresponds to the evolution of a tab of the \gls{cir} over slow time $t$.
	In the breathing case (i), the columns of $\tilde{\bm{R}}$ have approximately the shape of a line-of-sight component followed by a dense multipath profile. The magnitude of each column changes according to the periodic chest movement of the target. This is in agreement with the model of \cite{moederlEuRad2022,moederlICASSP2023}.
	In the talking case (ii), the chest motion is not periodic anymore and variations can be observed across different propagation delays.
	The maximum amplitude of the variations in the \gls{cir} ($\approx55\, \text{dB}$) is larger compared to the breathing case ($\approx 50\, \text{dB}$).
	In the movement case (iii), no clear structure is visible in the data and the maximum amplitude of the variations ($\approx 65\, \text{dB}$) is once again larger compared to the talking case. Note, that the colorscale in Figure \ref{fig:dataset:examples} is clipped such that all values $\geq 55\,\text{dB}$ map to dark blue.

	\section{Data-driven Occupancy Detection}
    \subsection{ResNet Architecture}
    We used two different neural-network architectures based on the ResNet architecture \cite{heCVPR2016:ResNet}. For each of these architectures, several variants were evaluated with a varying computational complexity.
    
\begin{figure*}
		\centering
		\hfill\subfloat[\label{fig:models:1d}]{\includegraphics[width=\columnwidth,valign=c]{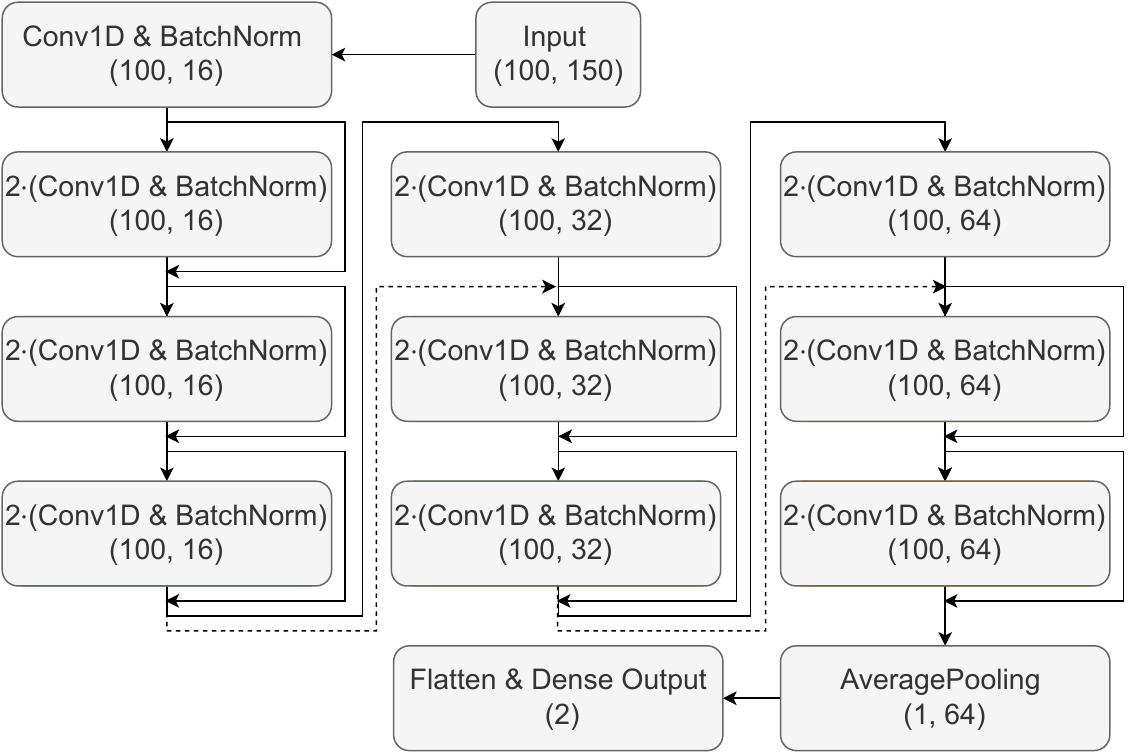}}\hfill\subfloat[\label{fig:models:2d}]{\includegraphics[width=\columnwidth,valign=c]{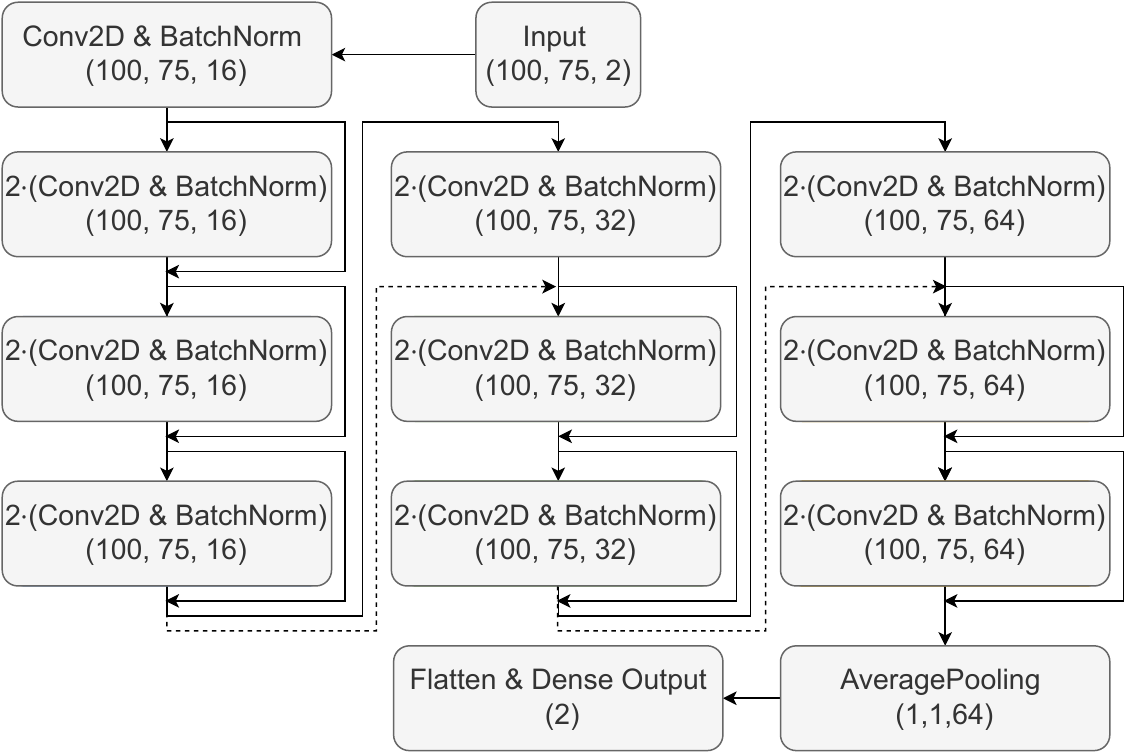}}\hfill
		\caption{Visualization of the ResNet-architectures for the (a) 1D-B and (b) 2D-A ResNet variants. For details of the architectures see Table \ref{tab:dl-architectures}.
		Connections and skip connections are denoted with arrows and projection skip connections are denoted with dotted arrows. The numbers in brackets denote the output dimensions of a block.}
		\label{fig:models}
	\end{figure*}

    The first architecture is a one-dimensional convolutional ResNet. Each sample $\tilde{\bm{R}}\in \mathbb{C}^{N\times M}$ is transformed to a sample $\bm{R_{\text{input}}}\in \mathbb{R}^{2N\times M}$ by stacking the real and imaginary parts on top of each other.
    The network architecture consists of a repeated structure of pairs of 1D-convolutional layers and batch normalization layers, with skip connections running in parallel to every pair of convolutional layers
    and batch normalization layers after the initial convolutions.
    We denote such a block as ``ResNet block''.
    After $N_{\text{double}}$ ResNet blocks, the number of convolutional filters (i.e. channels) doubles.
    A total of $N_{\text{total}}$ ResNet blocks is used. The 1D-convolutions are performed across the slow time dimension whereas the values in the fast time domain corresponded to the different channels of the input.
    To preserve the size across the slow time we used \emph{same padding}.
    The network ends with an average pooling layer across each channel, followed by a flattening and a dense output layer. This general architecture is shown in Figure \ref{fig:models:1d}.

    For the second architecture we used a two-dimensional convolutional ResNet. Each sample $\tilde{\bm{R}}\in \mathbb{C}^{N\times M}$ is transformed to a sample $\bm{R_{\text{input}}}\in \mathbb{R}^{N \times M \times 2}$ by adding a new dimension for the real- and complex-valued part of the sample. The network follows a similar structure as the one-dimensional ResNet.
    It consists of a repeated structure of 2D-convolutional layers and batch normalization layers, with skip connections running in parallel to every pair of convolutional layers.
    The convolutions are performed over the slow time and fast time dimensions while the last dimension corresponds to different channels.
    Like the 1D architecture the network ends with an average pooling layer across each channel followed by a flattening and a dense output layer.
    Figure \ref{fig:models:2d} illustrates this architecture.

    Both of these models were used with different variations in depth (number of layers) and width (number of filter maps), leading to different model complexities. Table \ref{tab:dl-architectures} lists the parameters for each model variant along with the number of trainable parameters and the \glspl{flop} of a single forward pass.
    
    \begin{table}
    	\centering
    	\caption{Evaluated model architectures.}
    	\label{tab:dl-architectures}
    	\begin{tabular}{r|ccccc}
    		\hline\hline
    		Variant & Initial filters & $N_{\text{double}}$ & $N_{\text{total}}$ & \# Parameters & FLOPs \\ \hline
    		1D-A & 32 & 3 & 12 & $1.5 \cdot 10^6$ & $3.0\cdot 10^8$ \\
    		1D-B & 16 & 3 & 9 & $1.0 \cdot 10^5$ & $2.0 \cdot 10^7$\\
    		1D-C & 16 & 2 & 6 & $6.8\cdot 10^4$ & $1.3 \cdot 10^7$ \\
    		1D-D & 16 & 1 & 3 & $3.5 \cdot 10^4$ & $6,9\cdot 10^6$\\
    		1D-E & 8 & 1 & 3 & $1.0\cdot 10^4$ & $2.1 \cdot 10^6$\\ \hline
    		2D-A & 16 & 3 & 9 & $2.7\cdot 10^5$ & $4.0 \cdot 10^9$ \\
    		2D-B & 16 & 2 & 6 & $1.7 \cdot 10^5$ & $2.6 \cdot 10^9$ \\
    		2D-C & 8 & 2 & 6 & $4.4\cdot 10^4$ & $6.5\cdot 10^8$ \\
    		2D-D & 8 & 2 & 4 & $1.1\cdot 10^4$ & $1.6\cdot 10^8$ \\
    		2D-E & 4 & 2 & 4 & $2.9\cdot 10^3$ & $4.1\cdot 10^7$\\
    		\hline \hline
    	\end{tabular}
    \end{table}

    \subsection{Data Augmentation and Training}
    Due to the high dynamic range and low noise floor of the laboratory equipment used in gathering the data, the detection process is trivial without data augmentation.
    However, in a real scenario the \gls{snr} of the \gls{uwb} radar will be lower compared to our laboratory equipment.
    Furthermore, only adults were measured in the dataset which have a large radar cross section and chest movement compared to infants. Thus, the expected variation in the \gls{cir} is much lower for an infant which further decreases the \gls{snr}.
    Let $\mathcal{R}_{\text{b}}$, $\mathcal{R}_{\text{t}}$, $\mathcal{R}_{\text{m}}$ and $\mathcal{R}_{\text{e}}$ denote the set of all samples of the breathing, talking and moving activities, respectively, as well as the samples from the empty car.
	Let the matrix operator $\|\cdot\|_2$ denote the Frobenius norm, such that $\|\tilde{\bm{R}}\|_2^2$ equals the energy of the received signal and $\bm{V}\in\mathbb{C}^{N\times M}$ denote a matrix of samples of the additive noise process $v(nT_{\text{ft}},mT_{\text{st}})$, sampled at $n\in \{0,1,\cdots,N-1\}$ and $m\in\{0,1,\cdots M-1\}$.
    To account for the lower \gls{snr} expected in a practical scenario, while also accounting for the different signal amplitudes at different activity levels, we define the $\text{SNR}=E_{\text{s}}/\|\bm{V}\|_2^2$ with respect to the median signal energy of the breathing case $E_{\text{s}} = \text{median}_{\tilde{\bm{R}}\in\mathcal{R}_{\text{b}}}(\|\tilde{\bm{R}}\|_2^2)$.
	During training, each sample of the occupied class (breathing, talking, moving) is used 200 times while each sample from the empty class is used 3000 times to account for the class imbalance. Each sample is augmented with noise such that an \gls{snr} drawn uniformly from the interval $[-30\,\text{dB},\,0\,\text{dB}]$ is achieved.
	Specifically, circularly-symmetric complex white Gaussian noise $\bm{V}$ with real and imaginary sample variance $\sigma^2 = E_{\text{s}}/(2MN\cdot 10^{\frac{\text{SNR}}{10}})$ is added to each sample instance $\tilde{\bm{R}}$ in $\mathcal{R}_{\text{b}}$, $\mathcal{R}_{\text{t}}$, $\mathcal{R}_{\text{m}}$ and $\mathcal{R}_{\text{e}}$, such that $\|\tilde{\bm{R}}\|_2^2 / \|\bm{V}\|_2^2$ equals the desired \gls{snr}.
	During the evaluation, we evaluate the test set once for each \gls{snr} $\in \{-10\,\text{dB},\,-11\,\text{dB},\,\cdots,\,-40\,\text{dB}\}$.
	Additionally, each augmented sample is normalized to unit energy before being fed into the ResNet.

    The data were split in training data and validation/test data between the different cars. For the training data, only the measurements from Car 1 are used. The data from Car 2 are used as validation and test data, with the last 150 samples of each class comprising the test set and the rest of the data from Car 2 being used for the validation set. While the validation set was used to govern early stopping during training, the test set was only used after the architectures have been designed and the networks have been trained.
    With this split, one of the two cars and 13 participants which were unseen during training are used only in the validation set and the test set.
    Note, that for the empty car, samples from Car 2 are available only.
    However, due to the added noise and the low energy of these samples, they are virtually indistinguishable from white noise after data augmentation anyways.
    Table \ref{tab:dataset-number-of-samples} lists the exact number of samples of each car used for training, validation and test.
    
   	\begin{table}
    	\centering
    	\caption{Number of Samples in the \datasetname\  dataset.}
    	\label{tab:dataset-number-of-samples}
    	\begin{tabular}{c|ccc|ccc}
    		\hline\hline
    		& \multicolumn{3}{c|}{Car 1 (Seat Leon)} & \multicolumn{3}{c}{Car 2 (Citro\"en C4 Picasso)} \\
    		& Train & Val. & Test & Train & Val. & Test \\\hline
    		Breathing & 368 & 144 & 0 & 0 & 409 & 150 \\
    		Talking & 367 & 145 & 0 & 0 & 406 & 150 \\
    		Moving & 380 & 161 & 0 & 0 & 410 & 150 \\
    		Empty & 0 & 0 & 0 & 66 & 100 & 20 \\
    		\hline\hline			
    	\end{tabular}
    \end{table}
 
	\section{Results}	
	\subsection{Classification Result}
\begin{figure}
		\centering
		\includegraphics{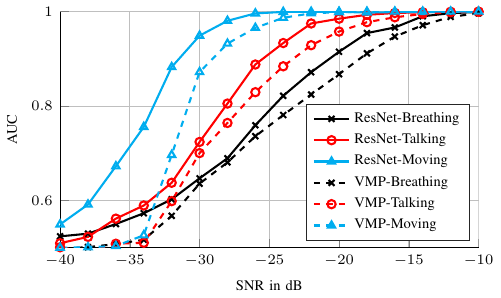}
		\caption{\acrshort{auc} over SNR for the 2D-A ResNet variant.}
		\label{fig:resutls:detection-performance}
	\end{figure}

	While the task of the ResNet is a binary decision between the car being empty or occupied, we evaluate the 3 different activities breathing, talking and moving separately due to the differences in scale and structure of the data.
	We measure the detection performance using the \acrfull{auc}, which is a threshold independent measure for a binary classifier.
	Figure \ref{fig:resutls:detection-performance} shows the \gls{auc} on the test data as a function of the \gls{snr} for the three different activities. The results are compared against a state-of-the art model-driven  approach based on \gls{vmp} \cite{moederlICASSP2023}, which was shown to outperform other model-driven approaches such as the estimate-correlator approach of \cite{moederlEuRad2022} or an \gls{fft}-based detector \cite{moederlICASSP2023}.
	As evident from Figure \ref{fig:resutls:detection-performance}, our ResNet is able to achieve a higher \gls{auc} than the \gls{vmp} detector in all three scenarios and over all tested \glspl{snr}.
	The performance difference is the highest for the movement scenario and somewhat smaller for the breathing and talking scenario.
	The assumptions of the model-driven approach are only (approximately) fulfilled in the breathing scenario.
	Nevertheless, even in the breathing scenario our ResNet-based approach outperforms the \gls{vmp} approach.
	We attribute this to model missmatch and the difficulty of specifying the correct prior for the physiological movement of the chest due to breathing in the \gls{vmp} approach.
	Because this biological process is hard to model, the \gls{vmp} approach assumes that the chest motion is a Gaussian process with a specified power spectral density.
	However, this prior is rather general.
	As indicated by the better detection performance, the ResNet seems to be able to (implicitly) estimate either a better fitting model or a more informative prior or both from the training data.
	The aforementioned model missmatch of the \gls{vmp} detector is higher for the stronger and more erratic movements of the talking and moving scenarios, which explains the higher performance difference between the ResNet and \gls{vmp} detectors in these cases.
		
	\subsection{Architecture Optimization}
\begin{figure}
		\centering
		\includegraphics{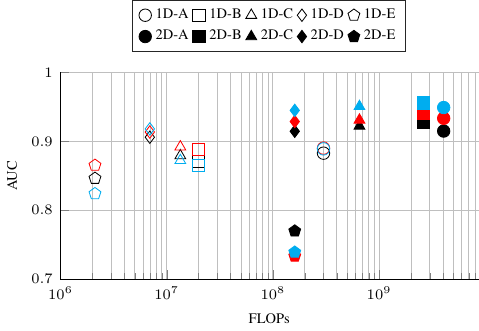}
		\caption{Detection performance of the different ResNet-architectures as function of the computational complexity of the forward-pass. Results are taken at $\text{SNR}=-20\,\text{dB}$ for the breathing activity (black), $\text{SNR}=-24\,\text{dB}$ for the talking activity (red) and $\text{SNR}=-30\,\text{dB}$ for the moving activity (cyan). Best viewed in color.}
		\label{fig:res:complexity}
	\end{figure}

	Since the detection algorithm is supposed to run on the resource-constraint onboard computer of a car, the computational efficiency of the ResNet forward pass is of interest.
	We performed an ablation study by gradually decreasing the number of channels, layers and filter banks in each layer of the ResNet for both the 1D and 2D variants.
	Figure \ref{fig:res:complexity} shows the resulting \gls{auc} values of the ResNets as a function of the number of \glspl{flop} required for the forward pass.
	Since the three different scenarios have different signal energies on average, we evaluate the \gls{auc} at $\text{SNR}=-20\,\text{dB}$ for the breathing subclass, at $\text{SNR}=-24\,\text{dB}$ for the talking subclass and at $\text{SNR}=-30\,\text{dB}$ for the moving subclass. The 2D-ResNet variants generally require more \glspl{flop} than the 1D-ResNet variants but achieve a slightly better performance. The best tradeoff between the computational complexity is achieved by the 1D-D ResNet variant which still outperforms the \gls{vmp}-based approach but requires less than $10^7$ \glspl{flop}. Note, that the \gls{vmp} approach is an iterative approach. Thus, the number of required \glspl{flop} is not constant and depends on the number of iterations needed to converge.
	
	\section{Conclusion}
	We present a ResNet-based car occupancy detection algorithm. The presented data-driven algorithm is shown to outperform state-of-the art model-driven occupancy detection algorithms, such as the \gls{vmp}-based approach of \cite{moederlICASSP2023}, in terms of \gls{auc} at low \glspl{snr} for all three activities breathing, talking and moving. In order to facilitate the implementation on the onboard computer of a car we evaluated several ResNet architectures to reduce the number of \glspl{flop} needed for a single forward pass of the network.
	The resulting 1D-ResNet architecture requires less than $10^7$ \glspl{flop} for a forward pass while still outperforming the \gls{vmp}-based approach.
	
	The \datasetname\ dataset \cite{moederl2023:Dataset} used to train and evaluate our detection algorithm is openly accessible to facilitate an easy and fair comparison between different approaches. This is in contrast to existing literature on \gls{uwb}-radar based car occupancy detection for which no open dataset exists to the author's knowledge.
	
	\bibliography{IEEEabrv,references_moederlRadarConf2023}
	\bibliographystyle{IEEEtran}
\end{document}